\documentclass[MSNbibl,nameyear,dvips]{arxstspdf}
\usepackage{flushend}
\usepackage{stfloats}
\volume{26}
\issue{1}
\pubyear{2011}
\firstpage{19}
\lastpage{20}
\doi{10.1214/11-STS337REJ}
\referstodoi{10.1214/10-STS337}

\begin{document}
\begin{frontmatter}

\title{Rejoinder}
\runtitle{Rejoinder}
\pdftitle{Rejoinder of Discussion of Statistical Inference: The Big Picture by R. E. Kass}

\begin{aug}
\author{\fnms{Robert E.} \snm{Kass}\corref{}\ead[label=e1]{kass@stat.cmu.edu}}

\runauthor{R. E. Kass}

\affiliation{Carnegie Mellon University}

\address{Robert E. Kass is Professor,
Department of Statistics, Center
  for the Neural Basis of
  Cognition, and Machine Learning Department, Carnegie Mellon
  University, Pittsburgh, Pennsylvania 15213, USA \printead{e1}.}

\end{aug}



\end{frontmatter}

In writing my essay I
presumed I was voicing, with a few novel nuances,
a nearly universal attitude among contemporary
statistical practitioners---at\break least among those who had
wrestled with the  incompatibility of Bayesian and frequentist logic.
Then David Madigan collected commentaries from several
thoughtful and accomplished statisticians.
Not only do I know
Andrew Gelman, Steve Goodman, Hal Stern and Rob McCulloch,
and respect them
deeply, but I would have been inclined to imagine I had been speaking
for them successfully. Their remarks shook me from my complacency.
While they generally agreed with much of what I had to say, there
were several points that would clearly benefit from additional
clarification and discussion, including the role of subjectivity in
Bayesian inference, the approximate
alignment of our theoretical and real worlds, and the utility of
$p$-values. Here I will ignore these specific
disagreements and comment further
only on the highest-level issues.

We care about our philosophy of statistics, first and foremost, because
statistical inference sheds light on
an important part of human existence, inductive reasoning,
and we want to understand it.
Philosophical perspectives are also
supposed to guide behavior, in research and in teaching.
My polemics focused on teaching, highlighting my discomfort with
the use of Figure~3 as the ``big picture'' of statistical inference.
My sense had been that as a principal
description of statistical thinking, Figure~3 was widely considered
bothersome, but no one had complained publicly. McCulloch agreed zealously.
Gelman and Stern, however, dissented; both find much
continuing use for the notion that statistics is largely about
reasoning from samples to populations. As a matter of classroom
effectiveness, I am sure that many instructors can do a great job
of conveying essential ideas of statistics using Figure~3. My main
point, though, was that introductory courses benefit
from emphasizing the abstraction of statistical models---their
hypothetical, contingent nature---along with the great utility of this
kind of abstraction. As we remarked in Brown and Kass (\citeyear{BroKas09}), when Box (\citeyear{Box})
said, ``All models are wrong, but some are useful,'' he was expressing
a quintessentially statistical attitude.  Figure~1  seeks
to make Box's sentiment central to statistical pedagogy, and I tried
to indicate the way the main idea may be illustrated repeatedly
throughout an elementary course.

Recognizing Box's  apparent influence here,
Goodman then asked whether I was simply restating Box's philosophy,
and he further prodded me to show how my own statement of
statistical pragmatism could be consequential.

In his 1976 Fisher Lecture, cited by Goodman,
Box railed against what he called ``mathematicity,'' meaning
theory developed in isolation from practice, and he stressed
the iterative nature of model building. The
fundamental role of
model criticism based on Fisherian logic was emphasized not only
by Box but also, in several roughly contemporaneous
discussions, by Dempster and by Rubin, and these presumably influenced
Gelman and Stern, who, together with their colleague Xiao-Li Meng,
developed and studied Bayesian model checking procedures.
Importantly, model criticism plays a prominent role
in Gelman et al. (\citeyear{Geletal04}).
The aim of my discussion, however, was somewhat different than what I
take Box to have intended.
I~understand Box to have said
that estimation should be Bayesian but criticism frequentist,
or inspired by frequentist logic. Statistical pragmatism
asserts, more simply and more generally,
that both forms of logic have merit, and either can be used
for any aspect of scientific inference.
In addition, I suggested the commonality of subjunctive statements
to help us acknowledge
that the big issues, in practice, are not Bayes
versus frequentist but rather the effects of various modeling
assumptions, and the likely behavior of procedures.

Stern noted that the pragmatism I described\break ``seems to be a fairly
evolved state for a statistician; it seems to require a clear
understanding of the various competing foundational arguments that
have preceded it historically.'' I agree. Along with Goodman, Stern
wondered whether such an eclectic philosophy could influence statistical
behavior, especially when tackling unsolved problems. I would claim
that it does. I admit, however, that
I have not done the substantial work it would take
to provide a satisfactory argument, with compelling examples. Lacking
this, I
will try to make do with a brief illustration.

Many experiments in neuroscience apply a fixed stimulus repeatedly to
some neural network and observe the consequences.
A typical example, discussed by Vu, Yu and Kass (\citeyear{VuYuKas09}), involved the
audio replay of a short snippet of a natural birdsong while a single
auditory neuron was recorded from a zebra finch. In such contexts,
mutual information is often used to
quantify the strength of the relationship between stimulus and response.
Mutual information requires the joint time series of
stimulus and response to be stationary and ergodic, but bird songs
contain many bursts of rapidly varying intensities with long pauses in
between. Thus, a snippet of natural song appears highly
nonstationary. In other experiments, the stimulus is
deterministic. Vu et al. asked whether, in such contexts,
estimates of mutual information become meaningless.
If we demand that there be a well-defined chance mechanism behind
every stochastic assumption, as the literal interpretation of
Figure~3 suggests, then clearly mutual information becomes void
for deterministic stimuli; but so too would any kind
of statistical inference involving the joint distribution.
The broader notion emphasized by Figure~1 is
that the mathematical formalism in the stochastic model is an
abstraction whose primary purpose is to represent, in all relevant respects,
the variability displayed by the data. Under this interpretation,
stochastic models can be of use even with deterministic stimuli.
Thus, dismissal of mutual information on the grounds of inadequate
chance mechanism is too crude. Instead, the constraint on
time series variability imposed by stationarity must be
considered carefully.
Vu et al. provided more pointed criticism,
some new mathematical
analysis, and a way to salvage the usual quantitative measures in such
settings. Was the philosophy behind Figure~1 necessary to obtain the
results of Vu et al.? No. But as I hope to have indicated, it
was helpful in supporting a path we could follow, and that is all one
should ask of foundations.\looseness=1

\vspace*{6pt}


\begin{thebibliography}{4}
\vspace*{6pt}
\bibitem[\protect\citeauthoryear{Brown and Kass}{2009}]{BroKas09}
\begin{barticle}[mr]
\bauthor{\bsnm{Brown},~\bfnm{Emery~N.}\binits{E.~N.}} \AND
  \bauthor{\bsnm{Kass},~\bfnm{Robert~E.}\binits{R.~E.}}
(\byear{2009}).
\btitle{What is statistics? (with discussion).}
\bjournal{Amer. Statist.}
\bvolume{63}
\bpages{105--123}.
\bid{doi={10.1198/tast.2009.0019}, issn={0003-1305}, mr={2750071}}
\end{barticle}
\endbibitem

\bibitem[\protect\citeauthoryear{Box}{1979}]{Box}
\begin{bmisc}[auto:STB|2011-03-03|12:04:44]
\bauthor{\bsnm{Box},~\bfnm{G.~E.~P.}\binits{G.~E.~P.}}
(\byear{1979}).
\bhowpublished{Robustness in the strategy of scientific model building.
  In \textit{Robustness in Statistics} (R. L. Launer and G. N. Wilkinson, eds.) 201--235.
  Academic Press, New York}.
\end{bmisc}
\endbibitem

\bibitem[\protect\citeauthoryear{Gelman et~al.}{2004}]{Geletal04}
\begin{bbook}[mr]
\bauthor{\bsnm{Gelman},~\bfnm{Andrew}\binits{A.}},
  \bauthor{\bsnm{Carlin},~\bfnm{John~B.}\binits{J.~B.}},
  \bauthor{\bsnm{Stern},~\bfnm{Hal~S.}\binits{H.~S.}} \AND
  \bauthor{\bsnm{Rubin},~\bfnm{Donald~B.}\binits{D.~B.}}
(\byear{2004}).
\btitle{Bayesian Data Analysis},
\bedition{2nd} ed.
\bpublisher{Chapman \& Hall/CRC,} \baddress{Boca Raton, FL}.
\bid{mr={2027492}}
\end{bbook}
\endbibitem

\bibitem[\protect\citeauthoryear{Vu, Yu and Kass}{2009}]{VuYuKas09}
\begin{barticle}[mr]
\bauthor{\bsnm{Vu},~\bfnm{Vincent~Q.}\binits{V.~Q.}},
  \bauthor{\bsnm{Yu},~\bfnm{Bin}\binits{B.}} \AND
  \bauthor{\bsnm{Kass},~\bfnm{Robert~E.}\binits{R.~E.}}
(\byear{2009}).
\btitle{Information in the nonstationary case}.
\bjournal{Neural Comput.}
\bvolume{21}
\bpages{688--703}.
\bid{doi={10.1162/neco.2008.01-08-700}, issn={0899-7667}, mr={2478315}}
\end{barticle}
\endbibitem

\end{thebibliography}
\end{document}